\documentclass[epj]{svjour}
\usepackage{graphicx}
\begin{document}
\title{Electron spin precession in semiconductor quantum wires
       with Rashba spin-orbit coupling}
\titlerunning{Spin precession semiconductor quantum wires\ldots}
\author{Manuel Val\'{\i}n-Rodr\'{\i}guez, 
Antonio Puente \and 
Lloren\c{c} Serra \thanks{E-mail: \email{DFSLSC4@clust.uib.es}}}
\authorrunning{M. Val\'{\i}n-Rodr\'{\i}guez {\em et al.}}
\institute{Departament de F\'{\i}sica,
Universitat de les Illes Balears, E-07122 Palma de Mallorca, Spain}
\date{April 16, 2003}
\abstract{
The influence of the Rashba spin-orbit coupling on the electron spin dynamics
is investigated for a ballistic semiconductor quantum wire with a finite width.
We monitor the spin evolution using the time-dependent Schr\"odinger equation.
The pure spin precession characteristic of the 1D limit is lost in a
2D wire with a finite lateral width. In general, the time evolution in 
the latter case is characterized by several frequencies and a nonrigid 
spin motion.}
\maketitle
\section{Introduction}

The control of the Rashba spin-orbit coupling is one of the most promising 
tools for the manipulation of the spin of the carriers within  
semiconductor heterostructures. In the last years, there have appeared
several  proposals of spin-based devices relying on this 
mechanism \cite{peet02,egue02}. 
Among them we cite the spin-FET, first proposed by  Datta and 
Das \cite{datt90}, the spin filters \cite{koga02} and the 
spin guides \cite{lolo02}.

The origin of the Rashba spin-orbit coupling in III-V semiconductor
heterostructures lies in their asymmetry. The lack of space symmetry causes 
a local electric field perpendicular to the plane of the heterostructure. 
The associated relativistic correction makes the electrons
feel an effective magnetic field which separates
in energy the different spin states \cite{ras60}.
This dependence of the Rashba spin-orbit coupling on the interface
electric field remains as a great advantage for spin 
control in semiconductors. Indeed, the intensity of the coupling can be 
controlled by applying a vertical electric field to the heterostructure,
as proved experimentally by Nitta {\it et al.} \cite{nitt97}. 

In this work, we shall focus on the spin precessional properties of
conduction electrons in a quantum wire
having a finite width. The problem of the quantum wire with Rashba
coupling has already been studied by 
Moroz and Barnes \cite{moro99}, who addressed the ballistic conductance
and the subband structure of the wire; and by 
Mireles and Kirczenow \cite{mire01}, who analyzed the spin-dependent 
transport within a tight binding model. 
Following a different approach,
Governale and Z\"ulicke \cite{gove02} have also treated the problem, and, 
more recently, Egues {\it et al} \cite{egue02b} have proposed to take advantadge
of the coupling of the first two transverse subbbands to create a spin-FET with
enhanced spin control. In this work, we shall concentrate on how the coupling
of the different transverse subbands, due to the Rashba term,
affects the electron spin precession.

The Hamiltonian representing the Rashba spin-orbit coupling in terms
of the electron momentum (${\bf p}$) and the Pauli matrices ($\sigma$'s) 
reads
\begin{equation}
\label{eq1}
{\cal H}_{SO} = 
\frac{\lambda_R}{\hbar} \left(\, p_y\sigma_x-p_x\sigma_y\,\right) \; ,
\end{equation}
where $\lambda_R$ represents the intensity of the spin-orbit coupling, 
depending on the heterostructure's vertical electric field. 
To stress the experimental feasibility we 
have assumed a value $\lambda_R=1.03 \times 10^{-9}$ eV cm,
as reported in Ref.\ \cite{nitt97} for an 
In$_{0.53}$Ga$_{0.47}$As/\-In$_{0.52}$Al$_{0.48}$As 
quantum well. 
Other experimental parameters corresponding to this sample
are: $m^*=0.05\, m_e$, $\epsilon=13.9$ for InGaAs, and the
Fermi wavevector $k_F=3.5\times 10^6$ cm$^{-1}$, 
corresponding to a 2DEG density $n_s\simeq 2\times 10^{12}$ cm$^{-2}$.
All calculations discussed below have been obtained using this parameter
set.

We assume a wire oriented along the $y$ axis with 
confinement achieved in the $x$ direction.
In the simplified model of a one-dimensional wire the terms containing 
$x$ or $p_x$ are dropped and the 
remaining Hamiltonian 
\begin{equation}
{\cal H}_{1D}=\frac{p_y^2}{2m^*}+\frac{\lambda_R}{\hbar}p_y\sigma_x
\end{equation}
is compatible with the observables $\{p_y,\sigma_x\}$. That is, 
the energy eigenstates are characterized by their $x$ spin component 
and $y$ momentum ($\hbar k_y$). 
In this system, the state of an electron having 
well-defined momentum and with arbitrary spin orientation decomposes
into a combination of the spin eigenstates $|s_x+\rangle$ and 
$|s_x-\rangle$, leading to a spin 
{\em precessional} motion about the $x$ axis, i.e., in the $S_y$-$S_z$ plane. 
The spin 
evolution is characterized by an angular frequency $\omega_P$, 
corresponding to the energy difference between the two spin eigenstates,
$\hbar \omega_P=2\lambda_R k_y$. 
This precession frequency leads
to the well-known formula for the angular modulation of spin orientation
$\theta_R=2m^*\lambda_R L/\hbar^2$, for a channel length $L$.

When a two-dimensional free
motion of the electrons is considered, the eigenstates of the Hamiltonian
\begin{equation}
{\cal H}_{2D}=\frac{\left(p_x^2+p_y^2\right)}{2m^*}+\frac{\lambda_R}{\hbar}
\left(p_y\sigma_x-p_x\sigma_y\right)
\end{equation}
have again well-defined momentum $\hbar{\bf k}$ and spin orientation, but 
the latter depends now on the wavevector ${\bf k}$.
Using spinorial notation in the usual basis of $|+\rangle$
and $|-\rangle$ states for the $S_z$ spin component, 
the wave functions and energies for a given
${\bf k}$ read
\begin{eqnarray}
\label{chi}
\chi_{{\bf k}\pm} &=&
e^{i{\bf k}\cdot{\bf r}}
\left(\begin{array}{c}
	1 \\
	\mp ie^{i\phi}
	\end{array}\right) \; ,\\
\varepsilon_{{\bf k}\pm} &=&
\frac{\hbar^2}{2m^*}(k_x^2+k_y^2)\pm\lambda_R\,|{\bf k}|\; ,
\end{eqnarray}
where $\phi={\rm atan}(k_y/k_x)$ is the polar angle for ${\bf k}$.
Note that the spinors (\ref{chi}) are eigenstates of a spin component 
orthogonal to ${\bf k}$, i.e.,  
the spin orientation depends on the 
spatial wavevector, always pointing in perpendicular 
direction to $\bf k$.
As in the 1D case, a state having
well-defined momentum $\hbar{\bf k}$ and arbitrary spin orientation
decomposes into a linear combination of the above 2D eigenspinors
$\chi_{{\bf k}\pm}$. 
Therefore, the evolution under ${\cal H}_{2D}$ leads again to a spin
precession of frequency $\omega_P$ given by the energy difference
between the corresponding 'up' and 'down' eigenstates.
Namely, $\hbar\omega_P=2\lambda_R|{\bf k}|$; the same result 
of the 1D case. An important difference is, however, that  
the plane of spin precession depends in 2D on the spatial orientation of 
${\bf k}$. Having analyzed the bulk limits, the question that arises is 
how the above {\em pure} spin precession is modified when the complete 
translational invariance of the homogeneous 1D and 2D systems is lost
in a wire with a finite width.

\section{The Model}

In order to model the quantum wire we maintain the translational
invariance in the longitudinal coordinate ($y$) while, in the transversal
direction ($x$) this symmetry is broken by a parabolic confinement
\begin{equation}
V(x)=\frac{1}{2}m^*\omega_0^2 x^2\; ,
\end{equation}
where $\omega_0$ determines the {\em width} of the wire. The Hamiltonian
representing the quantum wire reads
\begin{equation}
\label{eq7}
{\cal H} = \frac{{\bf p}^2}{2m^*}+
\frac{1}{2}m^*\omega_0^2\,x^2+
\frac{\lambda_R}{\hbar}\left(p_y\sigma_x-p_x\sigma_y\right)
\; .
\end{equation}

As our system still retains one spatial symmetry we can
reduce the spinorial 2D problem to a 1D one by factorizing out the
$y$ dependence,
\begin{equation}
\label{psi}
 \left(\begin{array}{c}
	\psi_{nk_y\uparrow}({\bf r}) \\
	\psi_{nk_y\downarrow}({\bf r}) 
	\end{array}
	\right) = e^{ik_yy}
 \left(\begin{array}{c}
	\phi_{nk_y\uparrow}(x) \\
	\phi_{nk_y\downarrow}(x)
	\end{array}\right)\; .
\end{equation}

The continuum index $k_y$ in Eq.\ (\ref{psi}) labels the state of
free motion in longitudinal direction while the discrete index $n$ 
accounts for the different transversal subbands of the confining 
dimension. 
Introducing this general form of energy eigenspinor we obtain the 
reduced Hamiltonian for the spinorial transverse dependence.
\begin{eqnarray}
\label{Ham}
{\cal H}_{\it tr} &=& \frac{\hbar^2 k_y^2}{2m^*}+\frac{p_x^2}{2m^*}+
\frac{m^*\omega_0^2}{2}x^2\nonumber\\
&+& \frac{\lambda_R}{\hbar}
\left(\hbar k_y\sigma_x-p_x\sigma_y\right)
\end{eqnarray}

The transverse Hamiltonian, Eq.\ (\ref{Ham}), can be analyzed by seeking
direct solutions to the stationary Schr\"odinger equation or,
equivalently, solving the time-dependent Schr\"od\-in\-ger equation
for given initial conditions. We shall use 
the latter approach since it yields a suitable scenario to address the 
dynamical spin evolution. For a given wavevector
$k_y$ the coupled time-dependent equations for the two
spin components of any spinorial wavepacket: $f_{k_y\uparrow}(x)$
and $f_{k_y\downarrow}(x)$ read
\begin{eqnarray}
&& i\hbar\partial_t
 \left(\begin{array}{c}
	f_{k_y\uparrow}(x,t) \\
	f_{k_y\downarrow}(x,t) 
	\end{array}
	\right) = \nonumber\\
\label{eq10}
&&  \left(\begin{array}{cc}
	{\cal H}_0^{(x)} & \lambda_R \left( k_y+\partial_x\right) \\
	\lambda_R \left( k_y-\partial_x\right) & {\cal H}_0^{(x)} 
	\end{array}
	\right)
 \left(\begin{array}{c}
	f_{k_y\uparrow}(x,t) \\
	f_{k_y\downarrow}(x,t) 
	\end{array}
	\right)\; ,
\end{eqnarray}
where
\begin{equation}
{\cal H}_0^{(x)}=-\frac{\hbar^2}{2m^*}\partial_x^2+\frac{m^*\omega_0^2}{2}x^2\; .
\end{equation}
Note that in ${\cal H}_0^{(x)}$ we have dropped the longitudinal kinetic
contribution, not important for time evolution. 

It is well known that in quantum mechanics time evolution can provide 
complete information about the stationary states of a Hamiltonian through 
spectral analysis. Indeed, for a time-independent Hamiltonian
an arbitrary wavepacket can be expanded as a series
in the stationary energy eigenstates 
\begin{equation}
\label{spinor}
 \left(\begin{array}{c}
	f_{k_y\uparrow}(x,t) \\
	f_{k_y\downarrow}(x,t) 
	\end{array}
	\right) = \sum_{n} A_{nk_y}\,e^{-\frac{i}{\hbar}\varepsilon_{nk_y}t}\,
 \left(\begin{array}{c}
	\phi_{nk_y\uparrow}(x) \\
	\phi_{nk_y\downarrow}(x)
	\end{array}\right) \; .
\end{equation}
Using the fast Fourier transform (FFT)
we can extract the harmonic frequencies ($\varepsilon_{nk_y}$ energies)
from the evolution of an arbitrary wavepacket.
Furthermore, if this analysis is repeated at different spatial points the 
space components of the energy eigenspinor $\phi_{nk_y\uparrow}(x)$
and $\phi_{nk_y\downarrow}(x)$, can also be obtained.

\section{Results}

The Hamiltonian of Eq.\ (\ref{eq7}), apart from kinetic and potential terms,
contains the longitudinal and transversal spin-orbit
contributions coupling $x$ and $y$ spatial motions with 
the electron spin.
It is worth noticing that there is an analytical limit to 
the solution of Eq.\ (\ref{eq10}) when
$p_x f_{k_y} \ll \hbar k_y f_{k_y}$. 
In this case the solutions are similar to those discussed above for
${\cal H}_{1D}$, i.e., 
the eigenstates are still plane waves in $y$ direction
with spin oriented along $x$ but now 
with an harmonic oscillator transverse profile. 
The corresponding energy spectrum
is also analytically known
\begin{equation}
\varepsilon_{nk_ys}^{(0)}=
\frac{\hbar^2k_y^2}{2m^*}+\left(n+\frac{1}{2}\right)\hbar\omega_0
+\lambda_R k_y s\; ,
\end{equation}
where $s=+1$ for $+x$ spin orientation (up) and $s=-1$ 
for $-x$ (down).  If we substract the longitudinal
kinetic term $\hbar^2k_y^2/2m^*$, common to both spin orientations, the
remaining band structure consists of doubly split transverse subbands,
linear in $k_y$ and having a common origin at $(n+1/2)\hbar\omega_0$.
Another analytical limit can be obtained neglecting $\lambda_R p_y\sigma_x$
in (\ref{eq7}); 
in this case, spin-orbit interaction only produces a little constant
shift in the energy levels that does not depend on spin orientation. 

As a check of our numerical procedure, Fig.\ 1 shows the results for a
simulation using the above simplified Hamiltonian
(neglecting $\lambda_R p_x\sigma_y$ in Eq.\ (\ref{eq7})). 
The parameters
used are: $\hbar\omega_0=3.5$ meV,
$k_y=1.0\times 10^6$ cm$^{-1}$ and the above mentioned spin-orbit intensity. 
The initial input
for the time-evolution is a Gaussian-shaped wavepacket having
spin oriented in $+z$,
\begin{equation}
\left(\begin{array}{c}
	f_{k_y\uparrow}(x,0) \\
	f_{k_y\downarrow}(x,0)
\end{array} \right)\equiv
\frac{1}{2\sigma\sqrt{\pi}}\,e^{\frac{-(x-x_0)^2}{2\sigma^2}}
\left(\begin{array}{c}
	1 \\
	0
\end{array} \right)\; .
\end{equation}
Using a finite $x_0$ and $\sigma\simeq \sqrt{\frac{\hbar}{m^*\omega_0}}$ we
ensure that the initial wavepacket is composed of several transversal
eigenstates in the low energy region. Figure 1 shows the spectrum corresponding
to the wavepacket's oscillations at an arbitrary $x$ point, 
as well as the transversal
densities extracted for the two lowest peaks. An excellent agreement between
numerical and analytical energies and eigenspinors is found. 
Figure 2 represents the dynamical spin evolution for the above initial spinor,
corresponding to a pure precession in excellent agreement with the analytical 
frequency $\hbar\omega_P=2\lambda_R k_y$.

Note that, in this analytical regime, spin precession is independent of the 
transversal state, as the energy gaps are
common to the different subbands. This statement implies that spin
evolution is independent of the transversal profile of the injected particles.
Numerically, we have checked that arbitrary initial profiles for the 
$x$ dependence lead to the same spin dynamics.

The above analytical regime no longer holds when the transverse motion
couples with the spin through the term $\lambda_R p_x\sigma_y$ in 
Eq.\ (\ref{eq7}).
Complete analytical solutions for the Hamiltonian including the full 
spin-orbit interaction are not available and numerical calculations are 
needed to solve the problem. 
The first question arising is how the band structure of
the simplified Hamiltonian is modified when transversal spin-orbit
coupling is included. Figure 3 shows the corresponding band structure for 
the same wire parameters considered above. In order
to clarify the effect of the transversal coupling the longitudinal kinetic term 
$\hbar^2 k_y^2/2m^*$ has been subtracted and the analytical energies
are shown with dotted lines.
It can be seen that for low values of $k_y$ there is good agreement between the simplified 
and full Hamiltonians; but this is not true when we approach 
a region of band crossings.
When the $(n+1)$-th down subband reaches the up branch of the
$n$-th subband, the transversal coupling produces a large effect, removing the band
degeneracy and opening a gap between these two 
branches, $(n+1,\downarrow)$ and $(n,\uparrow)$ \cite{moro99}.
The gap opened clearly depends on the transversal subbands involved; the higher transversal
state the larger the gap, as easily noticed from Fig.\ 3. 
It can also be observed that the range of  $k_y$'s where there is a good agreement 
between full calculations and analytical
model disminishes as the subband index increases. These changes on the level
structure suggest significant modifications of the spin precession 
due to the coupling with transversal motion.

As shown by Fig.\ 3, transversal coupling effects maximize at 
the analytical model crossings, occurring approximately when 
$2\lambda_R k_y=\hbar\omega_0$. The fulfillment of this condition establishes
a rough criterion to quantify the relevance of the transversal spin-orbit coupling.
Nevertheless, it must be also stressed that the coupling depends sensitively on the 
subband index $n$.

The influence of the transversal coupling can also be observed in the structure of
the eigenspinors. In the analytical model the eigenfunctions are states having
well-defined spin orientation. The transversal spin-orbit coupling mixes the different
subbands leading to a more complex structure of the eigenstates. Figure 4 represents
the densities corresponding to three lowest eigenstates in the region of maximum
band anticrossing, i.e., $k_y= 1.70\times 10^{6}$ cm$^{-1}$. The lowest energy state
corresponds to a branch without crossings and, therefore, its eigenvalue and eigenspinor
are very similar to those of the analytical model; spin density mainly oriented in $-x$ 
direction with a residual structure in $z$ orientation. 
The other two eigenspinors
differ substantially from those of the analytical model; spin has no predominant orientation 
and displays a complex texture very similar to 
those found in Ref.\ \cite{gove02}. Charge and spin densities are related to the 
eigenspinor components through the usual expressions
\begin{eqnarray}
\rho(x) &=&  \mid\!\phi_{nk_y\uparrow}(x)\!\mid^2
+\mid\!\phi_{nk_y\downarrow}(x)\!\mid^2 
\nonumber\\
m_z(x) &=& \mid\!\phi_{nk_y\uparrow}(x)\!\mid^2
-\mid\!\phi_{nk_y\downarrow}(x)\!\mid^2
\nonumber\\
m_x(x) &=& 2 {\rm Re}\left\{
\phi_{nk_y\uparrow}(x) \phi_{nk_y\downarrow}^*(x) \right\}\; .
\end{eqnarray}

Until now we have presented the effects produced by the full spin-orbit
coupling on the level scheme with respect to the analytical model, having
results in good agreement with those of Ref.\ \cite{moro99}.
The question that we address next is how this modified band structure 
reflects on the spin dynamics.
We consider first the case of a particle having well-defined spin orientation at 
initial time and $y$-momentum in the range of band anticrossings, i.e.,
an spinor as in (\ref{spinor}) evolving with   
$k_y= 1.70\times 10^{6}$ cm$^{-1}$ in Eq.\ (\ref{eq10}). The initial condition
thus reads
\begin{equation}
f_{k_y}(x,0)\equiv
\frac{1}{\pi^{1/4}}\left[\frac{m^*\omega_0}{\hbar}\right]^{1/4}\,e^{-\frac{m^*\omega_0}{2\hbar}x^2}
\left(\begin{array}{c}
	1 \\
	0
\end{array} \right)\; .
\end{equation}
The spatial part of this injected spinor is equal to that of the
lowest analytic subband. In this way, we ensure that it is
in the low energy range, very close to the first transversal subband.

The dynamical evolution of this injected particle is shown in Fig.\ 5. 
The spin trajectory on the $S_y$-$S_z$ plane is quite involved, manifesting a
multifrequential evolution.
Indeed, the Fourier transform of
the series shows that spin evolves in time as a superposition of several frequencies
corresponding to the energy transitions between the different anticrossed branches.
Since the gap opened by the transversal spin-orbit coupling varies from 
one subband to another, several precession frequencies are found in the
anticrossing region. On the contrary, the analytical model predicts
a common transition energy for all the transversal subbands (see Fig.\ 3),
leading to a unique precession frequency. This feature proves that spin evolution
in 2D wires noticeably depends on the transversal state of the particle.
In the particular case of Fig.\ 5 the precession
is characterized by two dominant frequencies, corresponding to the transitions
between the lowest branch and the next two higher branches, represented 
by dark-gray
arrows in Fig.\ 3. This two-frequency evolution yields a variable
spin amplitude, in agreement with the displayed trajectory.
The spin-orbit coupling makes the precessional frequencies depend
on $k_y$ in a non-trivial way. At the same time, it gives rise to a $k_y$-dependent
spin modulation angle that would lead to a certain degree of decoherence
in the spin transport through the wire.

Figure 6 shows the precessional
spectra corresponding to an initial spinor having well-defined spin 
(in $+z$ direction) and different transversal states. The precessional spectrum 
corresponding to the first transversal subband ($n=0$) is that represented in 
Fig.\ 5. 
The precessional spectra for the higher transversal states, $n=1$ and $n=2$, exhibit 
an enhanced multi-frequential character, since more transitions significantly 
contribute to the spin evolution.
It can also be seen that the strength spreads over a higher frequency interval 
as the subband index $n$ is increased. 
For low $k_y$'s, out of the anticrossing region, the spin evolution recovers a
single-frequency behaviour, with slight deviations from the analytical results
(cf.\ Fig.\ 2).
Note that, in general, a pure up ($+z$) spin injected into the wire 
will couple with the different spinor eigenstates and, therefore,
the multifrequential evolution can not be avoided. 
  
Another question of relevance is how the transversal size influences the precession.
From Fig.\ 4 we can estimate the width of the wire in $90-120$ nm for
$\hbar\omega_0=3.5$ meV. If we reduce the value of $\omega_0$ and maintain
the intensity of spin-orbit coupling the crossings of transversal bands  become more
frequent as subbands are less spaced. Figure 7 shows the corresponding band structure
in the case $\hbar\omega_0=0.70$ meV; from the densities of the eigenspinors
an estimated width of  $200-250$ nm can be obtained for this wire. Deviations 
from the analytical band structure are clearly visible in spite of the
reduced $k_y$ range displayed. This feature is reflected in the precessional spectra
which are fragmented even for relatively low $k_y$'s. 
By contrast, a wire of $\hbar\omega_0=14$ meV, having
an estimated width of $40-60$ nm, shows a band structure very close to that of the
analytical decoupled Hamiltonian, as displayed in Fig.\ 8. This also reflects
on the precessional spectra, which are single-moded and match the value given by 
the analytical formula $\hbar\omega_P=2\lambda_R k_y$.

In practical implementations of the Datta and Das spin transistor \cite{datt90}
the above discussed spin decoherence at subband crossings would 
strongly affect the device operation and, therefore, it should be 
carefully considered in the design of appropriate working parameters.
We can quantify the decoherence by representing the spin amplitude
after the first spin flip, i.e., after the first $\pi$ rotation.
Figure 9 represents this quantity as a function of wire width ($\omega_0$)
for a $k_y$ value in the anticrossing region, 
$\approx\hbar\omega_0/2\lambda_R$. It can be seen that as the wire
becomes wider (lower $\omega_0$), the multi-frequential precession
leads to an important reduction of the $z$ spin after the first 
inversion. 

\section{Summary}

In summary, we have studied the electronic spin precession in quantum wires
with Rashba spin-orbit coupling by means of real time simulations. 
A detailed 
study of the band structure for this system has been presented and compared
with a decoupled analytical model. It has been shown that deviations from the
simple spin precession regime are determined by the relative importance of
the longitudinal spin-orbit energy ($\lambda_R k_y$) and the transversal
confining energy ($\hbar\omega_0$) and that the spin precession fragments
into several frequencies whose weights depend on the transversal state of the 
particle. The dependence of the spin precession on the wire width
has also been studied and illustrated through the band structure.

This work was supported by Grant No.\ BFM2002-03241 
from DGI (Spain).


\begin{figure}[f]
\centerline{\includegraphics[width=3.25in,clip=]{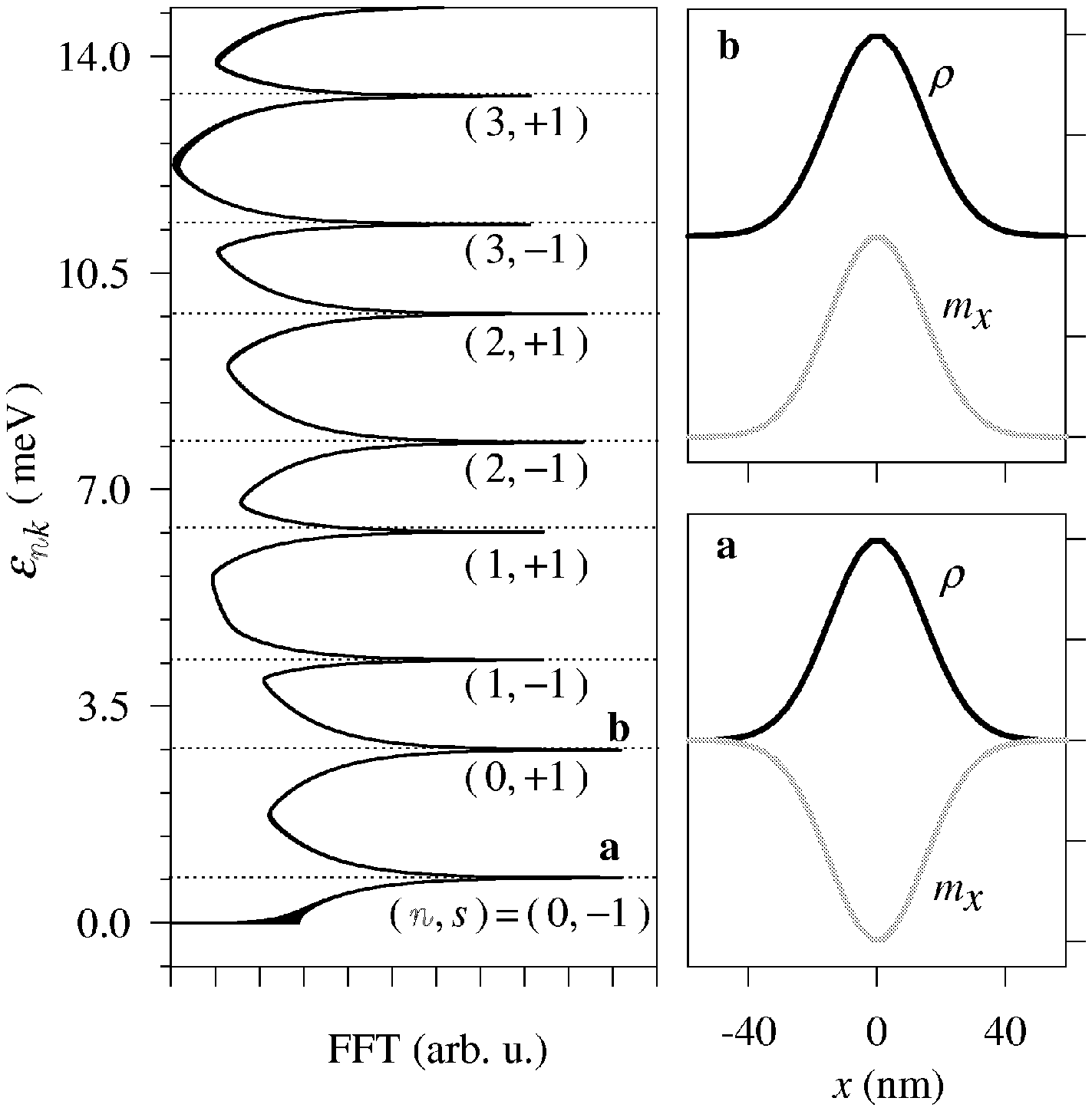}}
\caption{Fast Fourier transform (FFT) in logarithmic scale corresponding 
to the
oscillations of a {\em injected} wavepacket evolving with the simplified
Hamiltonian (without transversal spin-orbit coupling). 
Dotted lines indicate the analytical eigenvalues labeled by the 
transversal subband index $n$ and spin orentation $s$. 
Right panels show the non-zero densities corresponding to the two lowest 
eigenstates; charge and spin densities have been shifted in panel {\bf b} 
to better distinguish them.}
\end{figure}

\begin{figure}[f]
\centerline{\includegraphics[width=3.25in,clip=]{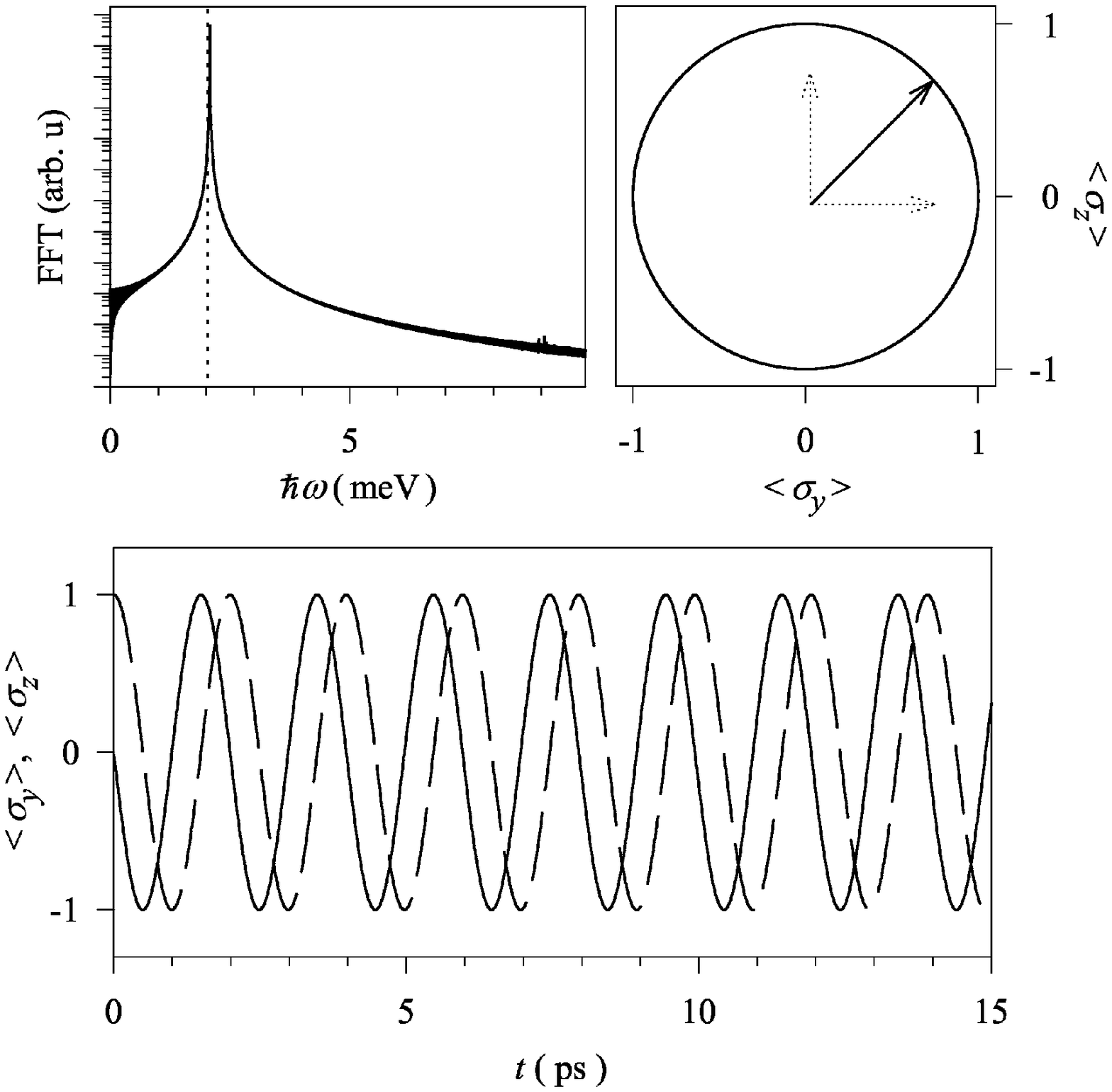}}
\caption{Lower panel: time-evolution of the spin expectation values in
the $S_y-S_z$ plane. Upper-left panel: Fourier
spectrum (logarithmic scale) corresponding to the $\langle\sigma_z\rangle(t)$
time series. The dotted line gives the analytical energy. 
Upper-right panel: trajectory in the $S_y-S_z$ plane of
the $\langle\vec{\sigma}\rangle$ vector.}
\end{figure}

\begin{figure}[f]
\centerline{\includegraphics[width=3.25in,clip=]{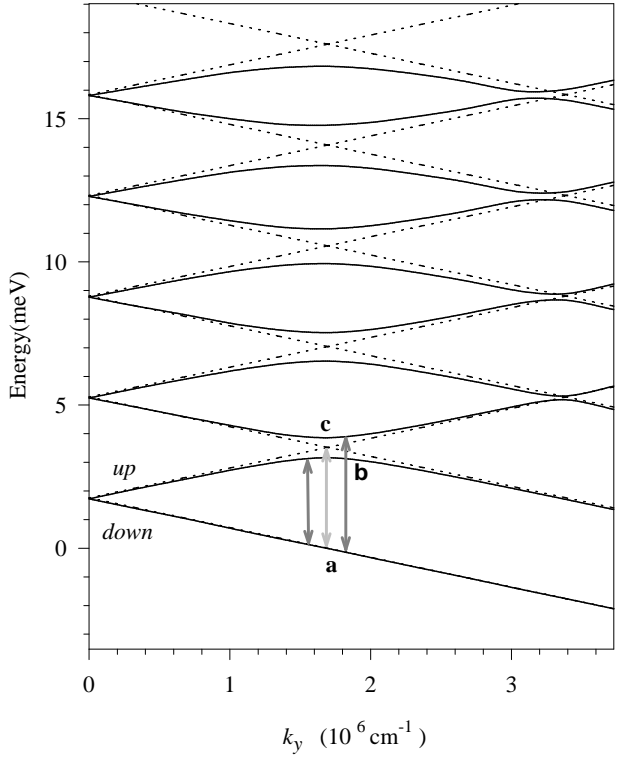}}
\caption{Band structure, excluding the constant kinetic term 
$\hbar^2k_y^2/2m^*$,
corresponding to the first five transversal subbands of a wire with 
$\hbar\omega_0=3.5$ meV.  Results for the full Hamiltonian (solid lines) and for 
the decoupled one (dotted lines) are shown. Transitions from the lowest down
branch in the decoupled (light grey) and full (dark grey) models
are indicated by arrows.}
\end{figure}

\begin{figure}[f]
\centerline{\includegraphics[width=3.25in,clip=]{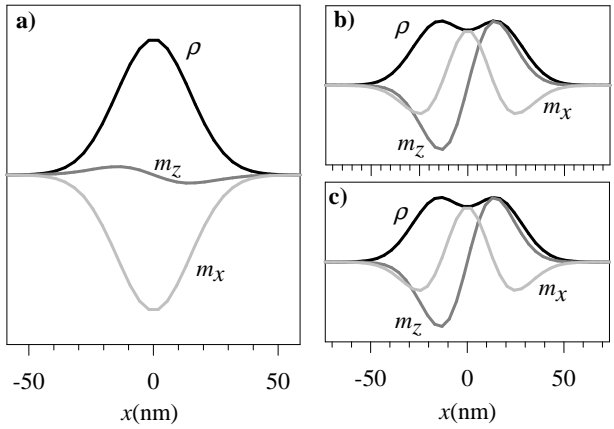}}
\caption{Charge and spin transversal profiles for the three lowest 
eigenstates of the full Hamiltonian in the anticrossing region.
The panel labels correspond to the 
{\bf a}, {\bf b} and {\bf c} branches indicated in Fig.\ 3.}
\end{figure}

\begin{figure}[f]
\centerline{\includegraphics[width=3.25in,clip=]{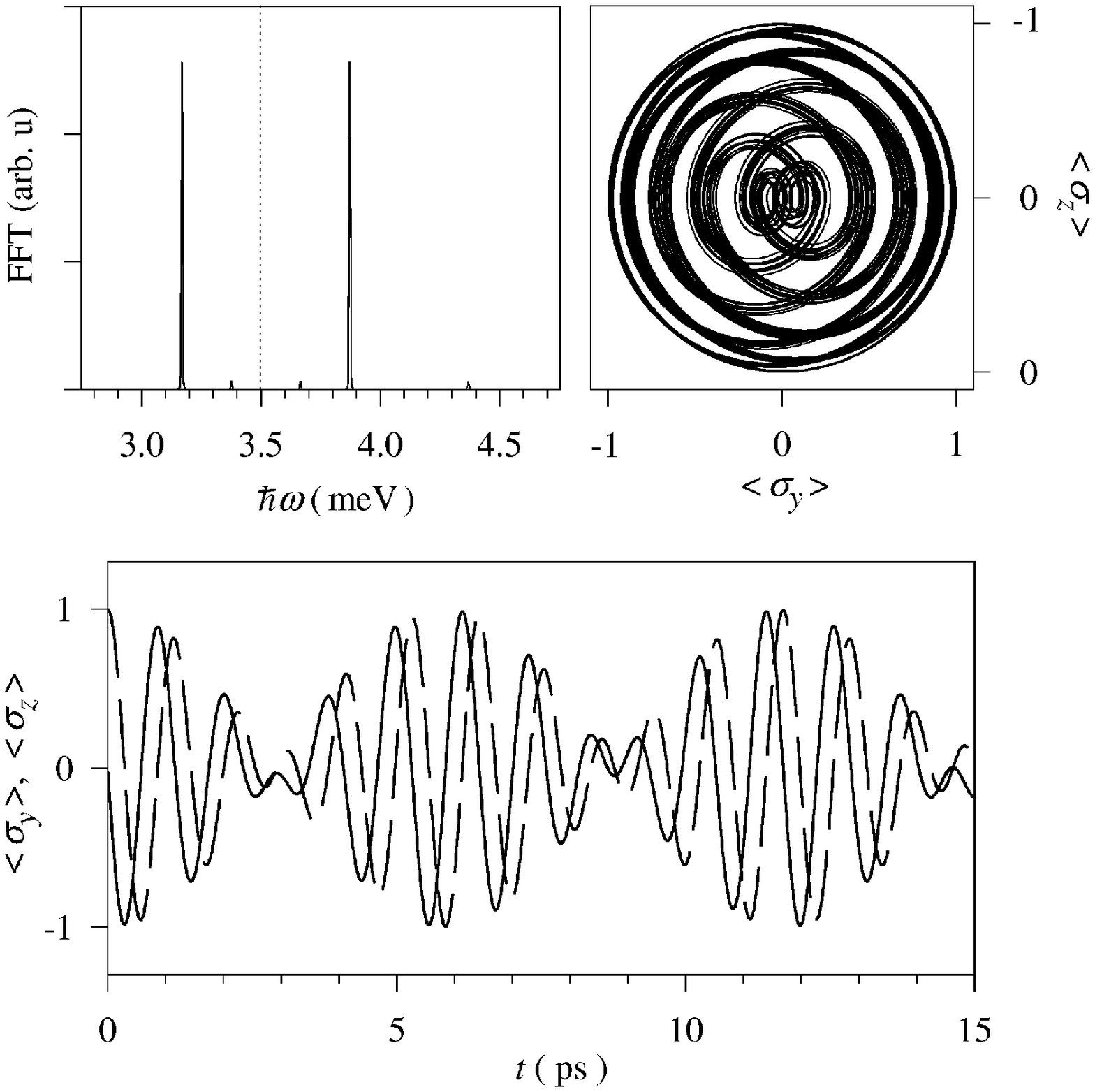}}
\caption{Lower panel: time-evolution of the $\langle\sigma_y\rangle$ and
$\langle\sigma_z\rangle$ expectation values for an injected wave-packet with
well-defined initial $z$ spin orientation using the full Hamiltonian. 
Upper-left panel: Fast Fourier transform (linear scale) of the 
$\langle\sigma_z\rangle(t)$ series; 
the dotted line indicates the precession frequency given by 
$\hbar\omega_P=2\lambda_R k_y$.
Upper-right panel: trajectory in the $S_y$-$S_z$ plane of 
the $\langle\vec{\sigma}\rangle$ vector.}
\end{figure}

\begin{figure}[f]
\centerline{\includegraphics[width=3.25in,clip=]{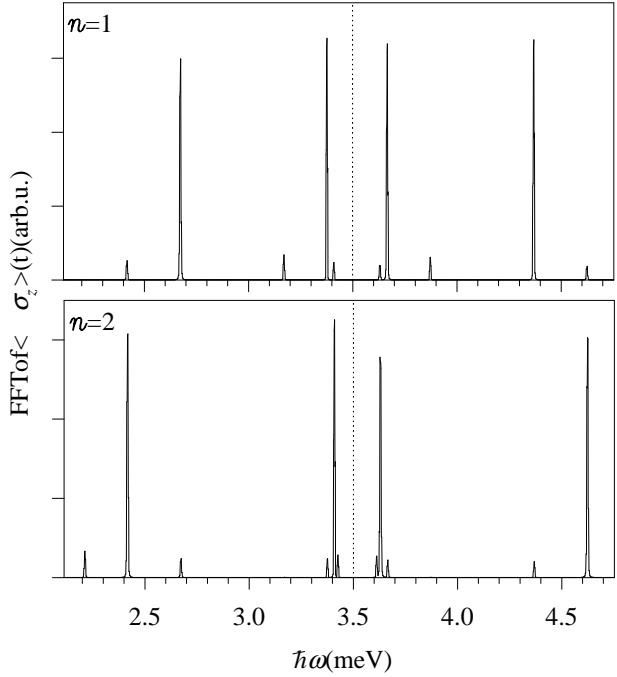}}
\caption{Precessional spectra obtained from the time-evolution of harmonic
spinors with well-defined $z$ spin corresponding to the $n=1$ and 2 
transversal subbands in the
anticrossing region. Dotted lines indicate the energy $\hbar\omega_P$.}
\end{figure}

\begin{figure}[f]
\centerline{\includegraphics[width=3.25in,clip=]{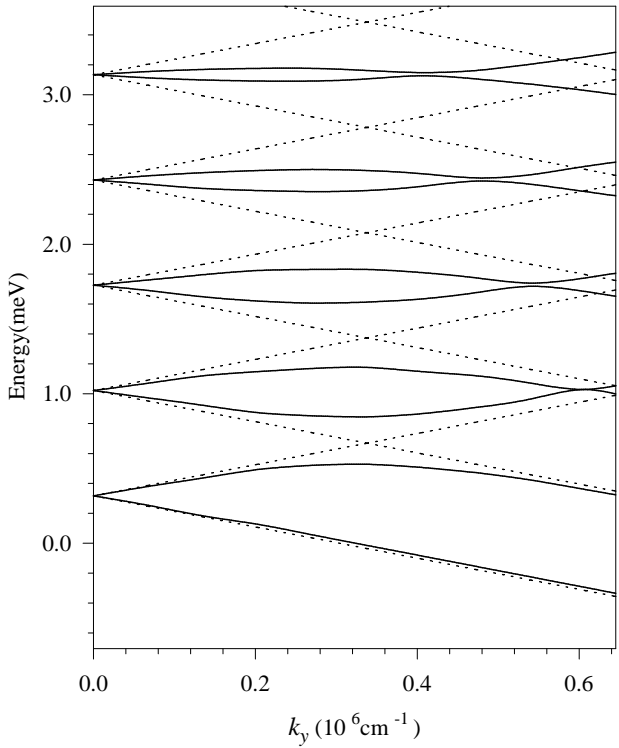}}
\caption{Band structure without the longitudinal kinetic term $\hbar^2k_y^2/2m^*$
corresponding to the first five transversal subbands of a wire with 
$\hbar\omega_0=0.7$ meV for the full Hamiltonian (solid lines) and for the 
decoupled one (dotted lines).}
\end{figure}

\begin{figure}[f]
\centerline{\includegraphics[width=3.25in,clip=]{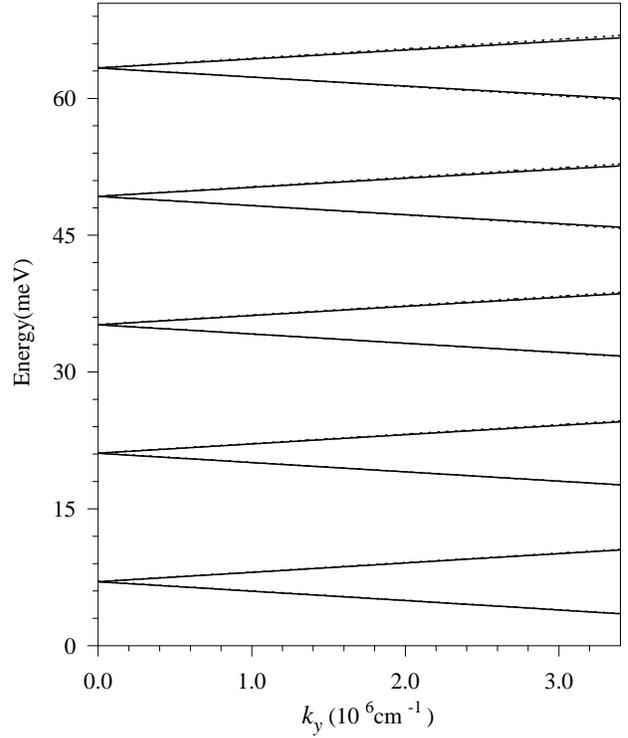}}
\caption{Same as Fig.\ 7 but for $\hbar\omega_0=14$ meV (Note the
enlarged $k_y$-scale with respect to Fig.\ 7).}
\end{figure}

\begin{figure}[f]
\centerline{\includegraphics[width=3.25in,clip=]{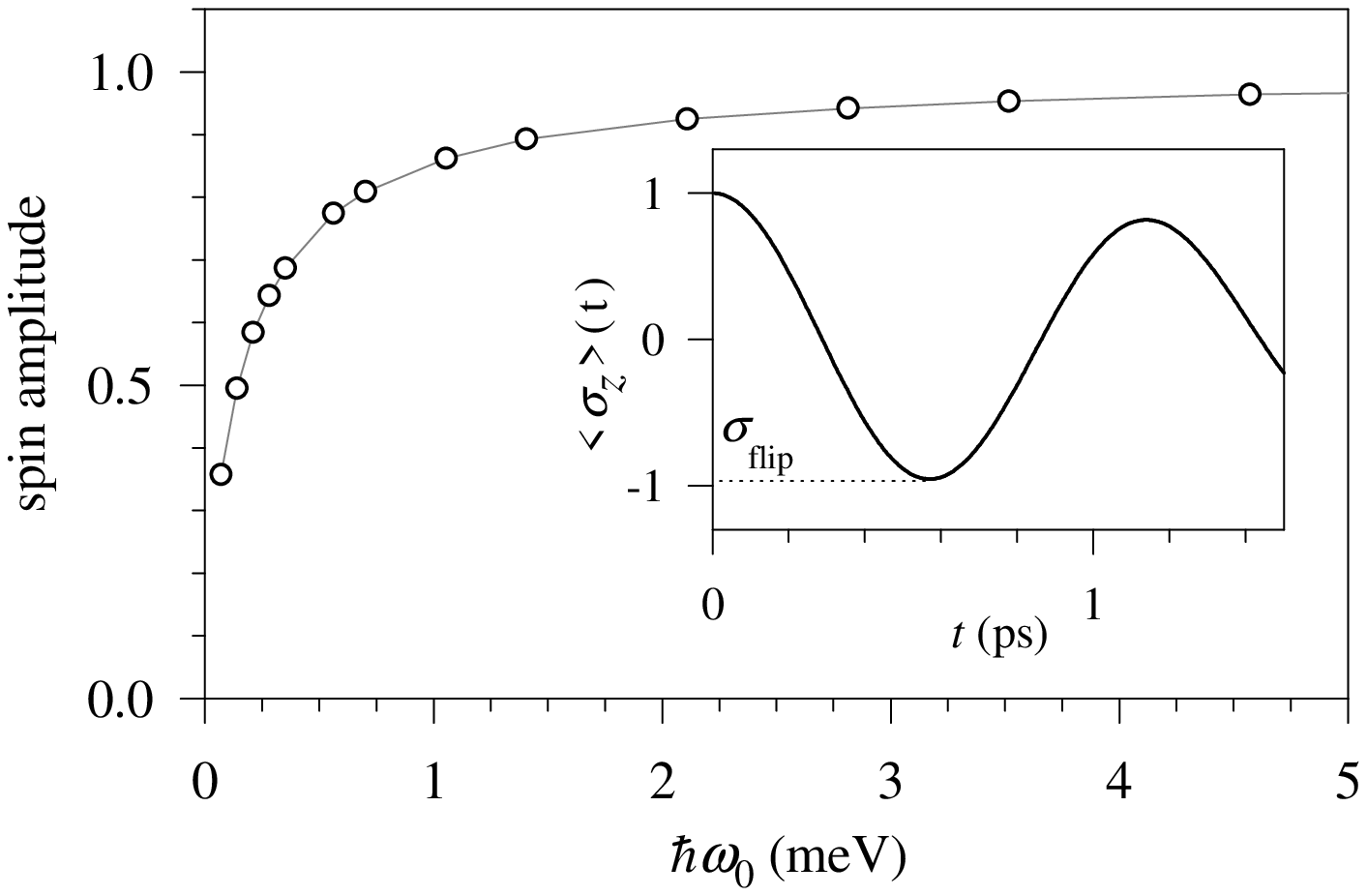}}
\caption{Amplitude of the $\langle\sigma_z\rangle(t)$ signal after 
the first spin flip as a function of the transversal confinement
strength $\hbar\omega_0$. The inset illustrates the displayed quantity
for the case $\hbar\omega_0=3.5$ meV.}
\end{figure}


\begin{thebibliography}{}

\bibitem{peet02}
X. F. Wang, P. Vasilopoulos, F. M. Peeters,
Appl.\ Phys.\ Lett.\ {\bf 80} 1400 (2002).

\bibitem{egue02} 
J. C. Egues, G. Burkard and D. Loss,
Phys.\ Rev.\ Lett.\ {\bf 89}, 176401 (2002).

\bibitem{datt90}
S. Datta and B. Das, 
Appl.\ Phys.\ Lett. {\bf 56}, 665 (1990).

\bibitem{koga02}
T. Koga, J. Nitta, H. Takayanagi, S. Datta,
Phys.\ Rev.\ Lett.\ {\bf 88}, 126601 (2002).

\bibitem{lolo02}
M. Val\'{\i}n-Rodr\'{\i}guez, A. Puente and Ll. Serra,
cond-mat/0211694.

\bibitem{ras60} E. I. Rashba, Fiz.\ tverd. Tela (Leningrad) {\bf 2},
1224 (1960) [Sov.\ Phys.\ Solid State  {\bf 2}, 1109 (1960)].

\bibitem{nitt97}
J. Nitta, T. Akazaki, H. Takayanagi and T. Enoki,
Phys.\ Rev.\ Lett. {\bf 78} 1335 (1997).

\bibitem{moro99} A. V. Moroz and C. H. W. Barnes,
Phys.\ Rev.\ B {\bf 60} 14272 (1999).

\bibitem{mire01} F. Mireles and G. Kirczenow,
Phys.\ Rev.\ B {\bf 64} 024426 (2001).

\bibitem{gove02}
M. Governale and U. Z\"ulicke, 
Phys.\ Rev.\ B {\bf 66}, 073311 (2002).

\bibitem{egue02b}
J. C. Egues, G. Burkard and D. Loss,
Appl.\ Phys.\ Lett.\ {\bf 82}, 2658 (2003).

\end{thebibliography}
\end{document}